\documentclass[prl,floatfix,amsmath,twocolumn,superscriptaddress,longbibliography]{revtex4-2}
\usepackage{latexsym}
\usepackage{amsmath}
\usepackage{amssymb}
\usepackage[ansinew]{inputenc}
\usepackage{graphicx}
\usepackage{txfonts}
\usepackage{booktabs}
\usepackage[colorlinks=true,linkcolor=blue,
pagecolor=blue,filecolor=blue,
menucolor=blue,urlcolor=blue,
citecolor=blue,anchorcolor=blue]{hyperref}
\newcommand{\GJ}{\textcolor{black}}
\newcommand{\JB}{\textcolor{black}}

\newcommand{\ZP}{\textcolor{black}}

\begin{document}

\title{Nonadiabatic braiding of Majorana modes}
\author{Fei Yu}
\affiliation{Center for Quantum Technologies, National University of Singapore, Singapore 117543}

\author{P. Z. Zhao}
\affiliation{Center for Quantum Technologies, National University of Singapore, Singapore 117543}

\author{Jiangbin Gong}
\email{phygj@nus.edu.sg}
\affiliation{Center for Quantum Technologies, National University of Singapore, Singapore 117543}
\affiliation{Department of Physics, National University of Singapore, Singapore 117551}

\begin{abstract}
The realization and manipulation of Majorana zero modes have drawn significant attention for their crucial role in enabling topological quantum computation. Conventional approaches to the braiding of \ZP{Majorana zero modes} rely on adiabatic processes. In this work, using a composite 2-Kitaev-chain system accommodating Majorana zero modes as a working example, we propose a \ZP{nonadiabatic and} non-Abelian geometry \ZP{phase}-based protocol to execute operations on these Majorana zero modes.  This is possible by locally coupling the edge sites of both quantum chains with an embedded lattice defect, successfully simulating the braiding operation of two Majorana modes \ZP{in a highly nonadiabatic fashion}. \ZP{To further enhance the robustness against control imperfections, we apply a multiple-pulse composite strategy to our quantum chain setting for  second-order protection of the braiding operations. Our proposal can also support the fast and robust realization of the $\pi/8$ gate, an essential ingredient for universal quantum computation.} This work hence offers a potential pathway towards the nonadiabatic and fault-tolerant control of Majorana zero modes.
\end{abstract}
\maketitle

\emph{Introduction} --- Localized edge modes and bulk-edge correspondence characterize distinct topological phases of matter \cite{schnyder2008classification,qi2011topological,kitaev2009periodic}. These edge modes, protected by system symmetry and bulk gap, are robust to weak local perturbations. Such topological robustness renders it ideal for fault-tolerant quantum information processing and storage. Kitaev first brings up the idea of topological quantum computation (TQC) using the one-dimensional (1d) \emph{p}-wave superconductor (SC) chain as a feasible platform \cite{kitaev2001unpaired,kitaev2003fault}. The edge modes in such \emph{p}-wave SC chains,  known as Majorana zero modes (MZMs) and exhibiting the non-Abelian anyonic statistics and non-local coherence, can be utilized to implement a quantum-gate-like operation by \JB{measurement-based fusion \cite{PhysRevLett.101.010501}} or mutual exchange (braiding). Since then, {great efforts} have been put into the realization of MZMs in fractional quantum Hall systems \cite{moore1991nonabelions}, 1d/2d SC heterostructures \cite{lutchyn2010majorana,sau2010generic,oreg2010helical}, and intrinsic 2d $p_{x} + ip_{y}$ SC \cite{sasaki2011topological,luke1998time,novak2013unusual}. \JB{In particular, the recent experimental breakthrough in indium arsenide-aluminium heterostructures with a gate defined
superconducting nanowire has paved a promising way towards fusion  operations on MZMs \cite{microsoft2025interferometric}}.

The existing braiding protocols of MZMs are often implemented adiabatically, \GJ{with one celebrated study}  reported in \cite{kraus2013braiding}. There, by manipulating the switch-on and switch-off of the local site-to-site coupling adiabatically, the operations can preserve the excitation \GJ{in the degenerate state manifold and maintain their nature as topological edge states}. An extended protocol under the same philosophy has been also implemented in nonequilibrium topological systems to implement the time-domain braiding of Majorana modes \cite{Raditya2018,PhysRevB.98.165421}. \GJ{From these studies, one common recognition is that nonadiabatic effects, arising from executing such protocols too rapidly, would detrimentally affect the braiding dynamics governed} by the Berry-Wilczek-Zee (BWZ) holonomy \cite{wilczek1984appearance,cheng2011nonadiabatic}.  The required adiabaticity, however, necessitates a much longer gate operation time, presenting challenges in practical implementations of TQC and making the system more vulnerable to control imperfections and decoherence \cite{schmidt2012decoherence,lai2020decoherence}. \GJ{To accelerate the operation protocols, one recently proposed route is to add the so-called counter-diabatic control terms to achieve shortcuts to adiabaticity} \cite{PhysRevB.91.201102,song2024shortcuts}. Besides, measurement-based method can also be utilized to speed up the braiding process and mitigate possible diabatic errors, as reported in ~\cite{PhysRevX.6.041003}. This work aims to design a robust nonadiabatic path toward fast operations on MZMs using concepts from holonomic quantum computation (HQC),  \JB{thus offering another digital scheme for operations on MZMs, complementary to measurement-based approaches \cite{PhysRevLett.101.010501,PhysRevX.6.041003,PhysRevB.101.085401,microsoft2025interferometric}.}

As an application of BWZ holonomy towards robust quantum computation \cite{zanardi1999holonomic},  HQC has been experimentally investigated on a number of platforms \cite{jones2000geometric,science.1058835,zu2014experimental}. The edge of HQC arises from the suppression of dynamical phases, such that the non-Abelian path-dependent geometric phase fully governs the qubit dynamics and hence the quantum gates are immune to a class of operational errors.  Indeed, the above-mentioned adiabatic manipulation of MZMs \cite{kraus2013braiding,Raditya2018,PhysRevB.98.165421} and other related studies \cite{karzig2016universal,karzig2019robust} can be regarded as HQC-based operations on Majorana modes but generalized to a many-body context. This work is inspired by the nonadiabatic version of HQC (NHQC) that inherits the main virtues from HQC  and further surpasses HQC by higher processing speed \cite{sjoqvist2012non,johansson2012robustness,xu2012nonadiabatic,wang2007nonadiabatic}.
Specifically, we propose, through NHQC protocols, to still harness the non-Abelian geometrical aspects of controlled quantum dynamics and achieve faster and fault-tolerant braiding operations on Majorana modes hosted by quantum chains.  Our innovation is twofold. First, we advocate applying local periodic driving to the end sites of quantum chains. This differs from the engineering of nonequilibrium topological phases where the periodic driving is often applied globally \cite{jiang2011majorana1, PhysRevB.87.201109}. Second, to facilitate the operations, we allow Majorana modes to temporarily leave the degenerate manifold by parking them as excitations in a single lattice defect. As shown below, in this way Majorana modes can also participate in the formation of useful dark states in the presence of driving, and local driving can safely return the Majorana modes to the quantum chain after acquiring a non-Abelian geometrical phase. Importantly, we show that, even in the presence of bulk states of the topological chain, there are no complications when we aim to further mitigate the impact of control errors by applying the multiple pulse composite-gate method in NHQC to our quantum chain setting.
\begin{figure}[tbp]
    \centering
    \includegraphics[width=1.0\linewidth]{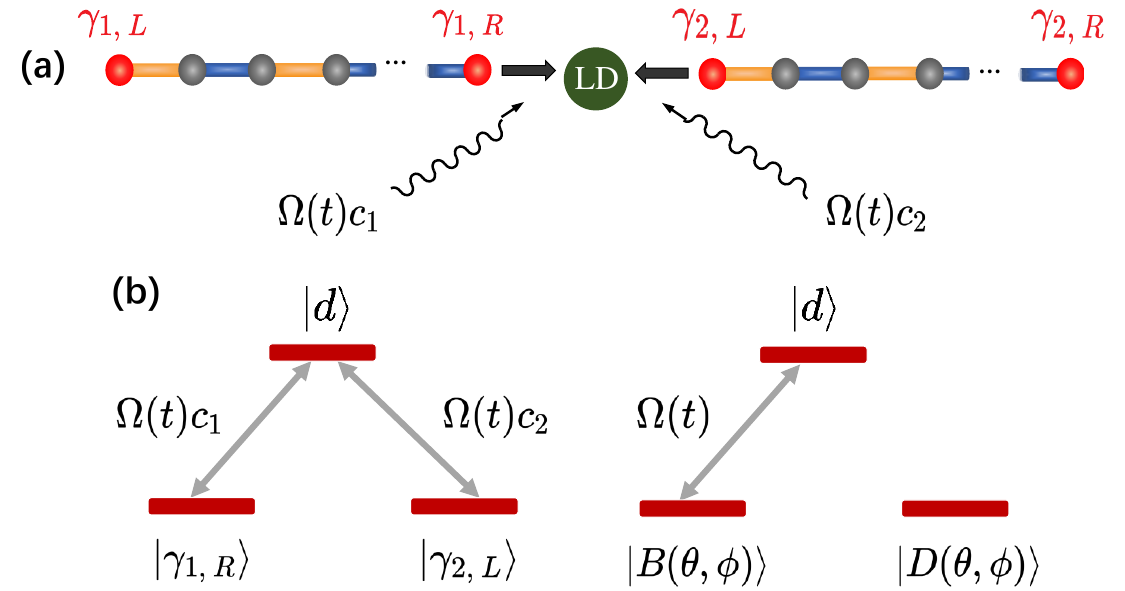}
    \caption{(a) Schematic diagram of the 2-quantum-chain configuration coupled with an inserted lattice defect (LD, green circle) through the local driving. (b) Illustration of the 3-level $\Lambda$-type Rabi oscillation occuring in our nonadiabatic braiding protocol. }
    \label{fig1}
\end{figure}

\emph{Model} --- We start with a composite atomic chain system as illustrated in Fig. 1(a), where both the first and second chains compose an $N$-site 1d $p$-wave SC (Kitaev) \GJ{lattice}. A lattice defect (LD, a single atomic site), or equivalently a quantum dot, is inserted between two chains and is initially \GJ{decoupled from both chains}. The initial static Hamiltonian $H_{\mathrm{stat}}$ can be written as follows:
\begin{equation} \label{eq1}
\begin{split}
H_{\mathrm{stat}} = \sum_{i = 1,\,2} \Bigg[&-\mu\sum_{n = 1}^{N} c_{n,\,i}^{\dagger}c_{n,\,i}-t\sum_{n = 1}^{N-1}\Big(\,c^{\dagger}_{n+1,\, i}c_{n,\, i}+\mathrm{H.c.}\Big)\\
&+\Delta \sum_{n = 1}^{N-1} \Big( c^{\dagger}_{n+1,\,i}c^{\dagger}_{n,\,i}+ \mathrm{H.c.}\Big)\Bigg]+\varepsilon_{d}\,d^{\dagger}d
\end{split}
\end{equation}
where $c^{\dagger}_{n,\,i}$\,($c_{n,\, i}$), $d^{\dagger}$\,($d$) are the fermionic operators on site $n$ of the $i$-th chain and on the lattice defect's site, respectively. $\varepsilon_{d}$ is the on-site energy of the LD, \GJ{whereas} $\mu, \, t, \, \Delta$ are the chemical potential, hopping amplitude, and SC pairing amplitude of the chains. Here we set $t = \Delta$ equally for two chains without loss of generality. When $|\mu|< 2t$, both chains are in the \GJ{topological nontrivial} phase.  As such, four localized MZMs $\gamma_{i,\, L/R}$ ($i=1,2$) emerge at the left/right sides of the $i$-th chain. In particular, the MZM at the right edge of the first chain is  $\left| \gamma_{1,\, R} \right>=\gamma_{1,\, R}\left|0\right>$,  and the MZM at the left edge of the second chain is $\left| \gamma_{2,\, L} \right>=\gamma_{2,\, L}\left|0\right>$, where $|0\rangle$ is the reference superconducting state of the SC chain excluding the LD's site. We also define the computational subspace $S$ spanned by $\left| \gamma_{1,\, R}\right>$ and $\left| \gamma_{2,\, L} \right>$.
Regardless of the specific values of the system parameters for the quantum chains, \GJ{many-body topological edge states} $\left| \gamma_{1,\, R}\right>$,  $\left| \gamma_{2,\, L} \right>$, and hence the computational subspace $S$ are pinned at zero energy. This invites a local periodic driving to couple the two MZMs in $S$ with the lattice defect state in a \ZP{controllable} fashion.

\GJ{We are now motivated to consider the following local driving that can actively introduce the crosstalk between $\gamma_{1,\,R}$ and $\gamma_{2,\,L}$, starting at time $t=0$}:
\begin{equation} \label{eq2}
H_{\mathrm{dri}}(t) = 2\Omega(t)\cos(\omega t+\phi_0) \left(c_1\, d^{\dagger}\gamma_{1,\,R} + c_2\, d^{\dagger}\gamma_{2,\,L} \right) + \mathrm{H.c.},
\end{equation}
where $\Omega(t)$ is the overall driving amplitude, \GJ{which is time-dependent from $t=0$ to the end of the protocol at $t=T$. The driving field frequency is set to be $\omega = \varepsilon_{d}/\hbar$, thus on resonance with the transition from MZMs to the lattice defect site. $\phi_0$ is a factor depicting the initial phase of the driving. To determine the explicit time dependence of the driving, there are two other time-independent coefficients $c_1$ and $c_2$ in $H_{\mathrm{dri}}(t)$.}

\GJ{A few remarks about the driving term $H_{\mathrm{dri}}(t)$ are in order.  (i) The wavefunctions of the two MZMs close to the defect are generally localized on very few sites. Therefore $H_{\mathrm{dri}}(t)$ only involves short-range pairing and hopping coupling with the LD. Such local driving is not unrealistic, considering available theoretical studies \cite{FTP2,FTP1} and experimental progress in the modulation of local coupling strength and external fields on the platforms of cold atom and quantum dot arrays \cite{PhysRevLett.111.047002,miles2025braiding1,RevModPhys.91.015005}; (ii) by applying the driving introduced above, one naturally expects to see a temporary lifting of the two MZMs out of the degenerate edge-state manifold.  To ensure that the MZMs can return to the topological chain at the end of the protocol, it is necessary to engineer the local driving, especially with the concern that once topological excitation is lifted, there will be issues regarding the dynamical phases.  As seen below, this problem can be resolved thanks to the physics of NHQC, based on which a braiding operation can be achieved rapidly.
(iii) Our investigations and simulations are always done with regard to the whole topological SC chain. It becomes curious to see how the local driving protocol can be digested based on MZMs and the LD, without explicitly referring to the SC chain.}

\emph{Braiding mechanism} --- Exploiting now that the lattice defect site is coherently coupled with both $\left| \gamma_{1,\, R} \right>$ and $\left| \gamma_{2,\, L} \right>$, one can identify a dark state decoupled from the local driving $H_{\rm dri}$. To see this, we denote  $\left|d\right>= d^{\dagger}\left|g\right>$, where $\left|g\right>$ is the unoccupied state on the defect's site.  Applying the rotating-wave approximation (RWA),  the $H_{\rm dri}$ in the associated rotating frame can then be easily rewritten as,
\begin{equation} \label{eq4}
H_{\rm dri}(t) = \Omega(t) \left(e^{i\phi_0}\left|d\right>\left<B(\theta,\,\phi)\right| + e^{-i\phi_0}\left|B(\theta,\phi)\right>\left<d\right|\right),
\end{equation}
where we have defined the bright state $\left|B(\theta,\,\phi)\right> = c_{1}\left|\gamma_{1,\, R}\right>+c_{2}\left|\gamma_{2,\, L}\right>$  \ZP{with $|c_{1}|^{2}+|c_{2}|^{2}=1$}.   The state $\left|D(\theta,\,\phi)\right> = c_{2}^{*}\left|\gamma_{1,\, R}\right>-c_{1}\left|\gamma_{2,\, L}\right>$,  which is also from the computational subspace $S$ but orthogonal to the bright state $\left|B(\theta,\,\phi)\right>$, does not even appear in $H_{\rm dri}(t)$ and is hence a dark state.  This useful physical picture is illustrated in Fig.~1(b).
At the end of the driving protocol at $t=T$, \ZP{if we require the pulse area of the driving satisfies 
$\int^{T}_{0}\Omega(t)dt =\pi$, one Rabi cycle between  state $\left|d\right>$ and state $\left|B(\theta,\,\phi)\right>$ is completed and then $\left|d\right>$ and  $\left|B(\theta,\,\phi)\right>$ acquire an Abelian $\pi$ geometric phase, while the dark state $|D\rangle$ is intact.} Therefore, assuming that the action of the rest of the topological lattice is essentially to park the concerned MZMs, the time evolution operator for $t=0$ to $t=T$ of the driven system is then given by
\begin{equation}\label{eq5}
U(T) =  -\left|d\right>\left<d\right|-\left|B\right>\left<B\right|+\left|D\right>\left<D\right|.
\end{equation}
\GJ{Because our initial state is from the computational subspace $S$, the quantum operation $U(T)$ returns all the population to the same computational subspace. More importantly, note that $H_{\rm dri}(t)$ at different times do commute (because only an overall factor changes with time), one can easily check that the matrix elements of $H_{\rm dri}$ between two arbitrary time-evolving state $\left|\psi_1(t)\right\rangle$ and $\left|\psi_2(t)\right\rangle$ emanating from the computational subspace $S$ is always zero, namely, $\langle\psi_1(t)\left|H_{\rm dri}\right|\psi_2(t)\rangle= 0$ at all times.  Therefore, there is no concern about imprinting different dynamical phases onto the two MZMs.
This makes it clear that, due to the joint action of the Abelian $\pi$ geometric phase and the intact dark state $\left|D(\theta,\,\phi)\right>$, $U(T)$ constitutes a nonadiabatic and non-Abelian geometric operation.
Indeed, projecting $U(T)$ on the computational subspace $S$, one obtains the following one-qubit holonomic gate}:
\begin{equation} \label{eq7}
U_{H}(\theta,\,\phi) = P_{S}U(T) P_S^{\dagger}=\mathbf{n} \cdot \boldsymbol{\sigma} = \begin{pmatrix}
\cos\theta  & \sin\theta \, e^{-i\phi} \\
\sin\theta \, e^{i\phi} & -\cos\theta
\end{pmatrix}
\end{equation}
\ZP{by setting $c_1=\cos(\theta/2)$ and $c_2=\sin(\theta/2)\exp(i\phi)$.} \JB{Note that the phase $\phi$ of $c_2$ can be introduced by a phase delay of the driving between the defect and the $\gamma_{2,\,L}$ mode from that between the defect and the $\gamma_{1,\,R}$ mode}.   Here, $\mathbf{n} = (\sin\theta\cos\phi,\,\sin\theta\sin\phi,\, \cos\theta)$ is a unit vector on the Bloch sphere, $\boldsymbol{\sigma} = (\sigma_x,\,\sigma_y,\,\sigma_z)$ are the Pauli operators, and $P_S = \left|\gamma_{1,\ R}\right>\left<\gamma_{1,\ R}\right|+\left|\gamma_{2,\ L}\right>\left<\gamma_{2,\ L}\right|$.  If we further set the parameters $\theta = \phi = \pi/2$, this holonomic gate would exactly become the braiding operator $U_{\mathrm{braid}} = \exp(\pi\,\gamma_{1,\,R}\gamma_{2,\,L}/4)$ for $\gamma_{1,\,R}$ and $\gamma_{2,\,L}$, up to a $U(1)$ phase.  This result represents the key idea behind our proposal.
\begin{figure}[tbp]
    \centering
    \includegraphics[width=1.0\linewidth]{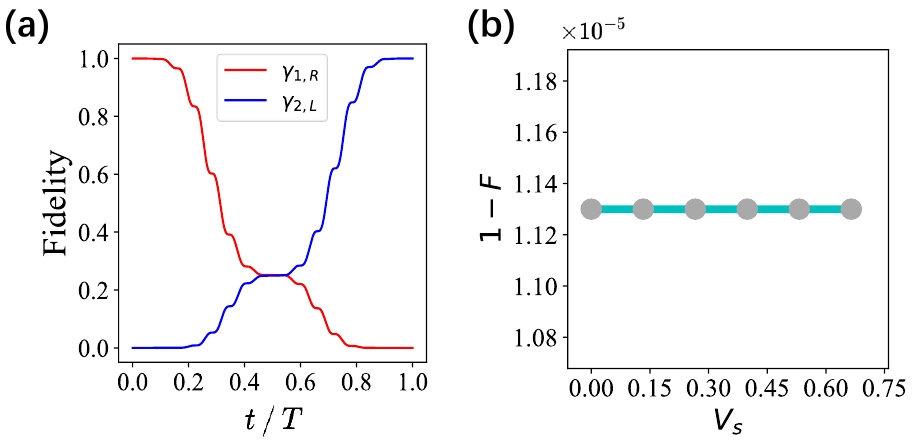}
    \caption{(a) Fidelity evolution of an initial $\gamma_{1,\,R} $ mode compared with itself (red curve) and the $\gamma_{2,\, L}$ state (blue curve) within one protocol for braiding. The parameters for calculation are $N=100$, $t = \Delta = 0.1$, $\varepsilon_{d} =3.0$, $\mu/t = 0.2$, $\theta = \phi = \pi/2$ and $\Omega(t) = \frac{\pi}{T} \sin^2(\frac{2\pi t}{T})$. In our simulations, we also set $T=20$ in dimensionless units. (b) Final fidelity loss $(1-F)$ as function of the local perturbation intensity $V_s$. The local perturbation term is introduced into both chains, which has the Gaussian form of $\Delta H=\sum_{n,i=1,2}V_s\exp[(n-n_0)^2/2{\sigma_0}^2] \,c^{\dagger}_{n,i}c_{n,i}$. Here we take $n_0 = 50$ and $\sigma_0=3$ for simulation and this result holds in general for other choices of $n_0$ and $\sigma_0$.}
    \label{fig2}
\end{figure}

\GJ{In our considerations above, one main restriction to the duration $T$ of the braiding protocol arises from the RWA we have adopted.  Specifically, given the driving frequency $\omega=\epsilon_d$, one needs to make sure that
$T\gg 2\pi/\epsilon_d$ to justify the RWA \cite{PhysRevA.88.054301, Pyvovar_2022}.
This is in contrast to adiabatic protocols \cite{kraus2013braiding,Raditya2018}, where we should at least have $T\gg 2\pi/\Delta_E$, with $\Delta_E$ being the quantum chain band gap in which the MZMs reside. For the parameters we use in Fig.~2, the band gap $\Delta_E$ is about twenty times smaller than $\epsilon_d$, indicating a much longer operation time scale for the implementation of adiabatic protocols than ours. Furthermore, the above reasoning of how an NHQC protocol achieves braiding of the two MZMs treats MZMs like isolated states detached from the quantum chain. This is of course an insightful but simplified picture.  It is necessary to perform numerical simulations to check if the main physics holds without RWA and in the presence of some local perturbations to the chain parameters. One typical computational example taking into account the entire quantum chain is presented in Fig.~2. In the shown example, the final fidelity of the braiding operation is nearly perfect, even though some local perturbation to the chain parameter $\mu$ is introduced into the system, with different intensity $V_s$. Such results indicate that our nonadiabatic braiding protocol can benefit and still preserve topological robustness against local perturbations. Moreover, so long as the defect energy scale is chosen to be sufficiently large, the duration of the control protocol is not a concern because the protocol is nonadiabatic by construction. }

\emph{Error mitigation} --- The success of our above protocol seems to necessitate the accurate control of all site-to-site driving coefficients as determined by Eq.~(2). For general cases where the MZMs may not be located at one site exclusively, it is almost impossible to perfectly predict the full profile of the unknown Majorana states and thus all the desired driving coefficients. In addition, some systematic errors in the driving always exist. Therefore, it is essential for us to analyze and further suppress the effect of imperfect control on our braiding scheme. Indeed, the control imperfections of interest in our protocol can be eventually classified into the mistake occurring on the bright state $\left|B'\right> = \left|B\right> + \left|\delta B\right>$, where $\left|\delta B\right>$ represents the deviation of an imperfectly estimated or implemented bright state $\left|B'\right>$ from the exact one. We first note that this deviation vector $\left|\delta B\right>$ can be expanded into the components on the quantum chain bulk states and the Majorana states $\left|\gamma_{1,\,R}\right>, \left|\gamma_{2,\,L}\right>$. That means the impact of control imperfections can always be attributed to two error-caused effects: (i) bulk state excitation and (ii) imperfections of the driving coefficients $c_1$ and $c_2$.
The first effect describes the potential error-induced coupling of the LD state with the bulk states of the two quantum sub-chains.  To avoid having the bulk state excited from the MZMs, one can set the energy of the defect, namely, $\varepsilon_{d}$ to be above the upper band top of the topological lattice. This way, only the lattice defect state, not the bulk states, is on resonance with the local driving. Thus, the bulk state contribution can be energetically suppressed. Computationally, we have performed a test and observed that a suppression of tens of times can be achieved if we move $\varepsilon_d$ from being inside the bulk to above the bulk top. Indeed,  numerical experiments shown in Fig.~2 already choose $\varepsilon_d$ this way.

After analyzing the effect of bulk excitation, we next focus on examining errors in the driving field parameters $c_1$ and $c_2$. Indeed, the effect of error-distorted driving coefficients $c_i\to(1+\epsilon_i)\,c_i$ ($i=1,2$) is equivalent to shifting the driving parameters $\Omega(t)\to \Omega'(t)$ and $\theta \to \theta'$ in Eq.~(3) simultaneously.  For such systematic errors, the time evolution operator of our above proposal would be affected to the first order of errors. 
\begin{figure}[tbp]
    \centering
    \includegraphics[width=1.0\linewidth]{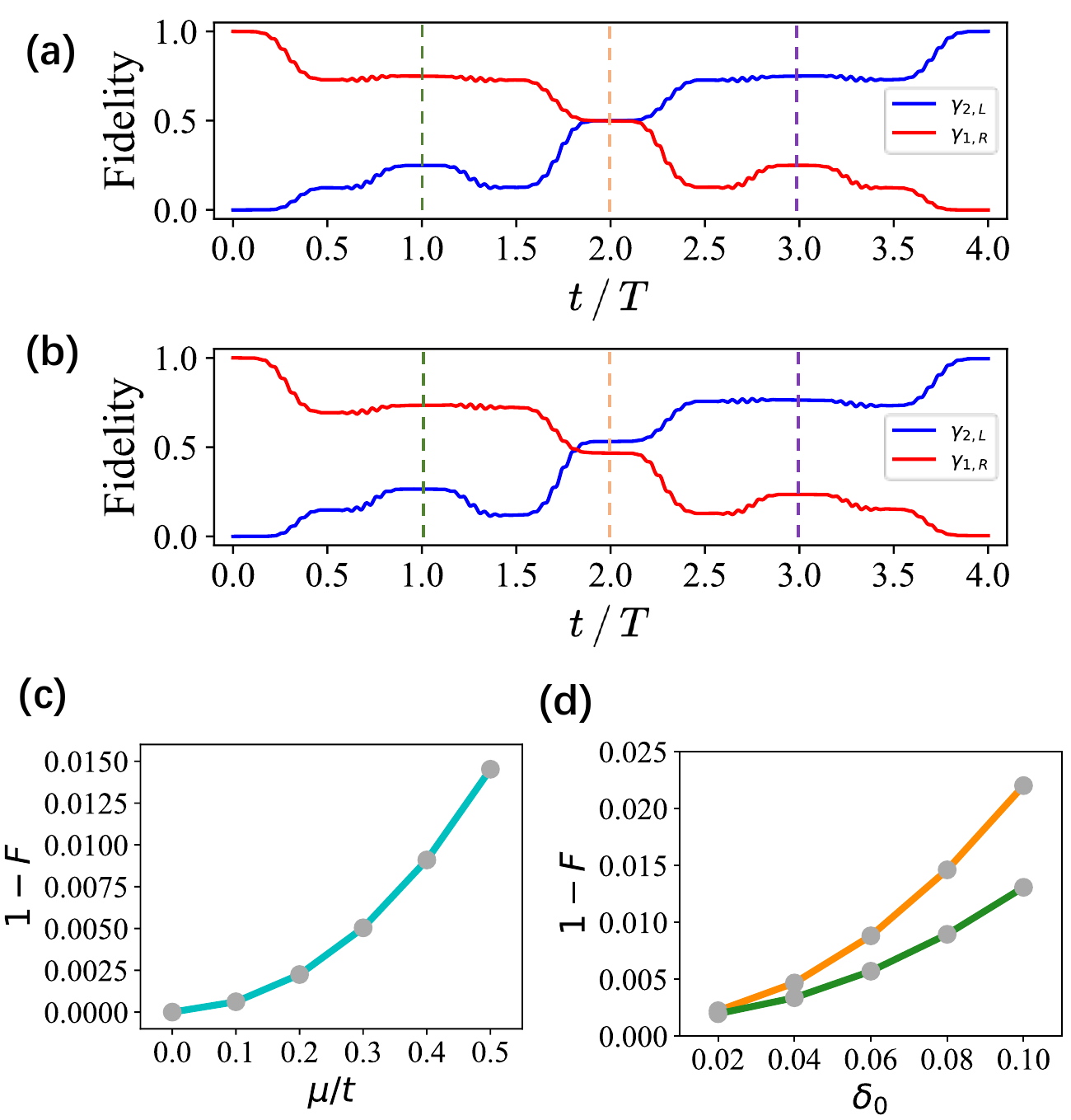}
    \caption{(a) Braiding fidelity evolution of an initial $\gamma_{1,\,R} $ mode with the application of composite gate method. During the period of $0\sim T$ ($2T\sim 3T$) and $T\sim 2T$ ($3T\sim 4T$), the segmented gate $U_{S}(\theta, \phi)$ ($U_{S}(\pi-\theta, \phi)$) is implemented. Here we choose the gate coefficients as $\theta=\pi/4$, $\phi=0$ and $T'=T/2$. Other parameters for calculation are the same as in Fig.~2(a). (b) Same as (a) but in the presence of random local coupling errors with control error strength $\delta_{0} = 0.06$, averaged over 100 times of implementation. The final fidelity loss is about $0.6\%$.  (c) Final fidelity loss for braiding $1-F$ vs. the chemical potential $\mu$ after implementing one-site truncation. (d) $1-F$ vs. the driving error strength $\delta_0$, with (green curve) and without (orange) the usage of composite gate method. \GJ{Other system parameters are the same as in Fig.~2.} }
    \label{fig3}
\end{figure}
To enhance the robustness against these driving coefficient errors, we further investigate whether a previous composite-gate method for NHQC proposed in Ref.~\cite{xu2017composite} can be applied to our quantum chain setting. The basic idea of such a method is to use the same local driving $H_{\rm dri}(t)$, but reconstruct the NHQC evolution path by multiple holonomic gate segments, with each segment satisfying the following driving parameter conditions:
\begin{equation}\label{eq11}
\int_{0}^{T'} \Omega(t) \, dt = \int_{T'}^{T} \Omega(t) \, dt = \frac{\pi}{2},\;\;\phi_0 = \varphi - \frac{\pi}{2} \, \Theta(t- T')
\end{equation}
where $t\in[0,T]$ is the evolution time of each segment, $T'$ is an arbitrary time between $0$ and $T$, and $\Theta(t)$ is the Heaviside step function. Under this design, the evolution of each segment is divided into two parts with different initial phases $\phi_0$, with the first (second) part of the evolution being from $t=0$ ($t=T'$) to $t=T'$ ($t=T$). Combining these two parts together, one can find that the time evolution associated with the whole segment yields the holonomic gate $U_{S}(\theta, \phi) = e^{i\,\pi/4}e^{-i\,\pi\, \mathbf{n} \cdot \boldsymbol{\sigma}/4}$ in the computational space $S$.
Next, by applying the holonomic gate segments $U_S$ multiple times, the accumulated errors in different segments can compensate for each other, assuming all segments suffer from the same type of errors.  Specifically, by applying the gate segment twice with the same parameters $\theta$ and $\phi$, one can cancel out the effect of amplitude errors $\Omega'$ to the second order at the level of the time evolution operator \cite{Single-loop}, whereas the $\theta'$ errors can be suppressed by applying a gate segment $U_{S}(\theta,\, \phi)$ followed by another segment $U_{S}(\pi - \theta,\, \phi)$ \cite{xu2017composite}.  Combining these two ideas, one can construct the following 4-segment composite gate to correct the $\epsilon_1$ and $\epsilon_2$ errors simultaneously:
\begin{equation}\label{eq12}
\begin{split}
U'_{\mathrm{comp}} &= U_{S}'(\theta, \phi)U_{S}'(\theta, \phi)U_{S}'(\pi-\theta, \phi)U_{S}'(\pi-\theta, \phi)\\
&= \exp[i(\pi-2\theta) \, \mathbf{v} \cdot \boldsymbol{\sigma}] + O(\epsilon_1^2) + O(\epsilon_2^2)
\end{split}
\end{equation}
where $\mathbf{v} = (-\sin\phi,\, \cos\phi,\, 0)$ is the tangent unit vector on the equator and the prime symbol denotes the gates affected by errors modeled by $\epsilon_{1}$ and $\epsilon_{2}$. Therefore, at the level of the time evolution operator,  the first-order errors in both $\epsilon_1$ and $\epsilon_2$ can be suppressed. Here one should set the gate coefficients $\theta=\pi/4$ and $\phi=0$ for the execution of a braiding process.

In Fig.~3 (a)-(d), we computationally evaluate the error resilience offered by the composite gate method, explicitly taking into account all lattice sites of our composite quantum chain. Since the Majorana states and thus the driving are highly localized, we expect that the local coupling $H_{\mathrm{dri}}$ between the defect and the subchains can be truncated within a very few sites' distance from the LD. Such truncation of driving can also be viewed as one type of control imperfection. In Fig.~3(c), we plot the braiding fidelity loss after one-site driving truncation as the function of chain parameter $\mu$. It is shown that when $\mu$ is small ($\mu<0.5$) so that the MZMs are relatively localized, the error from truncation can be mitigated to a comparatively low level by our NHQC design. This allows our system to further deviate from the ideal coupling in the form of $d^{\dagger}\gamma_{1,\,R}$ or $d^{\dagger}\gamma_{2,\,L}$. It is seen from simulation results in Fig.~3(b) that the final braiding fidelity curve, with random errors introduced but suppressed by the composite gate method, does not deviate much from the curve in Fig.~3(a) as the ideal case without any error. Since we have verified that the one-site truncation will not bring significant changes to the braiding fidelity, here we only consider the control errors in site-to-site driving amplitudes between the LD and its nearest neighbor sites as uniformly random variables in the range $[\delta_0, -\delta_0]$.  In Fig.~3(d), we quantitatively compare the control error resilience level with or without the use of the composite gate scheme. It is obvious that the composite gate method is necessary when errors quantified by $\delta_0$ are considerable.

\emph{Discussion} --- The MZMs belong to the Ising type of anyons characterized by the fusion rule $\sigma \times \sigma = 1 + \psi$ and $\sigma \times \psi = \sigma$. Braiding operators of such anyons form a complete set of the Clifford gates, but unfortunately, not the universal quantum gates. To support universal TQC on the MZMs platform, one should at least be capable of producing the "magic" state $\left|\psi\right>=e^{-i\pi/8} \left|0\right> + e^{i\pi/8} \left|1\right>$ in the Majorana system, or equivalently, performing the $\pi/8$ gate operation on the computational space $S$  \cite{bravyi2005universal}. However, this $\pi/8$ gate is not topologically protected or fault-tolerant, since it does not form an entire evolution loop in the degenerate state manifold. In this case, other conventional error-correction schemes in quantum computing would be required. For instance, the idea of adiabatic HQC combined with universal dynamical decoupling (UDD) is employed in Ref.~\cite{karzig2016universal} to construct a robust $\pi/8$ gate. Our proposal in this work can offer an alternative possibility to achieve the $\pi/8$ gate.  Indeed, our nonadiabatic gate operation can use a shorter protocol to escape decoherence, where its holonomic nature can also yield robustness against systematic errors.  To that end, we only need to modify the driving parameters $\theta, \phi$ defined above for the $\pi/8$ gate case ($\theta=3\pi/8,\, \phi =0$), and the composite gate method mentioned earlier can still be applied and be effective in mitigating the errors. 

Having confirmed that braiding operations can be done by tentatively lifting MZMs from the degenerate edge state manifold, this work will stimulate other interesting quantum control scenarios to be applied to topological edge states.  As an interesting extension, one may take some special bulk states to play the role of an intermediary \JB{between two MZMs localized at the same edge}. In that case, by adding local driving, one may directly couple MZMs with some bulk states, and with some necessary engineering, just as the lattice defect case considered here, to yield nontrivial operations on MZMs.  To enhance the coupling between MZMs and some particular bulk states, one may add an onsite quasiperiodic potential to the topological lattice to localize the bulk states in the vicinity of the MZMs.  This way, it is possible, \GJ{as confirmed by our computational simulation} (not shown here),  to execute the whole nonadiabatic braiding process on a single chain. 
In addition, our proposed approach can be applied to nonequilibrium topological lattices, e.g., a periodically driven (Floquet) Kitaev chain \cite{jiang2011majorana1, PhysRevB.87.201109}, to realize the fast braiding of an MZM and one Majorana $\pi$ mode \cite{ZHU20222145}. One may also extend our protocol to directly execute nontrivial unitary operations on more than two MZMs.   

\emph{Conclusion} --- In conclusion, we have proposed how to make use of a nonadiabatic holonomic quantum computation protocol to implement fast braiding of Majorana modes, under the 2-Kitaev-chain setting with a lattice defect and local driving between the chain and the defect. The underlying physics of nonadiabatic holonomic quantum computation can be clearly understood by isolating the MZMs from the lattice. By taking into account the whole topological lattice, the lattice defect plus the local driving,  it is shown that the proposed braiding protocol is feasible and that the composite gate method to mitigate the influence of driving parameter errors is effective. Our nonadiabatic protocol can also be used to implement the missing $\pi/8$ gate for universal TQC.

\emph{Acknowledgement} --- This project is supported by the National Research Foundation, Singapore through the National Quantum Office, hosted in A*STAR, under its Centre for Quantum Technologies Funding Initiative (S24Q2d0009).  J.G. thanks Gil Refael for valuable discussions during his visit to Singapore.  We are also grateful to Raditya Bomantara for valuable feedback on the first version of this manuscript. 

\bibliography{refs}

\end{document}